\documentclass[preprint,aps]{revtex4}
\usepackage{graphicx}
\usepackage{dcolumn}
\usepackage{bm}
\usepackage{subfigure}

\usepackage{amssymb}

\usepackage{verbatim}

\begin{document}

\title{Light Trapping in Thin Film Disordered Nanohole Patterns: Effects of Oblique Incidence and Intrinsic Absorption}

\author{Minhan Lou$^1$, Hua Bao$^{1}$\footnote[1]{To whom all correspondence should be addressed. Email: hua.bao@sjtu.edu.cn}, and Changying Zhao$^2$,}

\address{$^1$University of Michigan - Shanghai Jiao Tong University Joint Institute,\\
$^2$School of Mechanical Engineering,\\
Shanghai Jiao Tong University, Shanghai, China, 200240}

\begin{abstract}
Finite-difference time-domain method is employed to investigate the optical properties of semiconductor thin films patterned with circular holes. The presence of holes enhances the coupling of the incident plane wave with the thin film and greatly enhances the absorption performance. For a typical 100 nm thin film, the optimal hole pattern is achieved when the hole radius is 180 nm and volume fraction is about 30\%. Disorderness can alter the absorption spectra and has an impact on the broadband absorption performance. The non-uniform radius of holes can slightly broaden the absorption peaks and enhance the integrated absorption. Random hole position can completely change the shape of the absorption spectra and the averaged integrated absorption efficiency is slightly smaller than the optimized ordered nanohole pattern. Compared to random positioned nanoholes or ordered nanohole, amorphous arrangement of nanoholes will result in a much better absorption performance. However, it is also found that the absorption enhancement of amorphous pattern over an ordered pattern is weak when the incident angle departures from normal or when the intrinsic material absorption is strong.
\end{abstract}

\maketitle

\section{Introduction}

Thin film solar cells can reduce the cost of raw material and be assembled on flexible substrate, and thus have attracted significant research interest recently\cite{Agrawal_2008, Zeng_2006, Mallick_2012, Catchpole_2008, Ferry_2010}. However, due to the extremely small thickness (hundreds of nanometers in general) of these thin films, the optical absorption is quite small, which strongly limits the energy conversion efficiency\cite{Agrawal_2008}. To achieve high absorption in thin film, many photon management schemes have been proposed to enhance light trapping in semiconductor thin film, such as one-dimensional grating\cite{Zeng_2006}, photonic crystal \cite{Mallick_2012}, plasmonics\cite{Catchpole_2008, Ferry_2010}, random textured surface \cite{Rockstuhl_2008}, textured transparent electrode \cite{Muller_2004}, nanowires or nanoholes \cite{Tsakalakos_2007, Hu_2007, Han_2010}, etc. Theoretical investigations showed that for structures with feature size comparable to the wavelength, the light absorption enhancement can exceed the conventional Yablonovitch 4$n^2$ limit \cite{Yu_2010}. This provides large space for optical design at nanoscale to achieve further light trapping in these thin films.

Most of the theoretical investigations have been focused  on ordered structures, for example, on nanowire or nanohole arrays with square or triangular lattice  \cite{Han_2010, Lin_2009, Li_2009a, Bao_2010b}. However, by considering a thin film with shallow grating, Yu \emph{et al.} presented a theoretical analysis based on statistical temporal coupled mode theory and show that aperiodicity could be helpful to enhance broadband absorption, because symmetry can sometimes prevent the guided resonances in the thin film structure from coupling with outside \cite{Yu_2010b}. Bao \emph{et al.} demonstrated based on numerical electromagnetic simulations that in vertically aligned nanowire arrays with random position, diameter, and length, there is further enhanced light absorption compared to their ordered counterparts \cite{Bao_2010c}. Many numerical simulations have also been carried out to optimize aperiodic structure to further enhance the ultimate absorption efficiency and the results are quite encouraging \cite{Lin_2012,  Muskens_2008}. For example, Sturmberg \emph{et al.} numerically demonstrated that for an array of nanowires with optimized diameter distributions, the ultimate absorption efficiency can be 28\% larger than an array with uniform diameter \cite{Sturmberg_2012}. Although numerical results are demonstrated, the mechanism of absorption enhancement in disordered nanowire or nanoholes has not been fully understood. The different explanations include enhanced multiple scattering \cite{Bao_2010c} or better matching of solar spectra using leaky mode resonance \cite{Cao_2009, Sturmberg_2012}. Recently, Vynck \emph{et al.} studied the a two-dimensional slab with randomly positioned hole pattern and pointed out that amorphous pattern with short range correlation could achieve better broadband absorption enhancement \cite{Vynck_2012} in a relatively broad frequency range. On the other hand, most of the disordered structures have been compared to their ordered counterparts, which are generally not optimized. Normal incidence was usually considered in most of these investigations \cite{Bao_2010c, Lin_2012, Vynck_2012}, which could provide misleading conclusion because the solar cells will not operate at normal incidence in most cases. Also, the solar spectrum has been considered when evaluating the performance of solar absorption. Though it is necessary from an application point of view, it is hard to decouple the contribution of intrinsic material absorption and the enhancement due to the nanopattern.

In this manuscript we systematically investigate the light trapping in thin semiconductor film with disordered nanoholes using finite-difference time-domain (FDTD) simulations. The effects of random nanohole position and non-uniform radius are discussed. We also discuss the importance of oblique incidence and intrinsic material absorption on the broadband absorption performance of ordered and disordered patterns.

\section{Simulation details}

\begin{figure}[t]
\centerline{\includegraphics[width=0.4\textwidth]{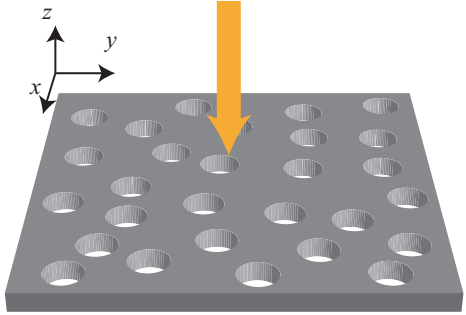}} \caption{\small The semiconductor slab with circular holes we investigate in this work.  }
\vspace{3mm}\label{Illustration}
\end{figure}

In this work, we consider a square semiconductor slab containing circular holes to mimic the ultrathin film solar cell, as shown in Fig. \ref{Illustration}. Periodic boundary condition is applied in the $x$ and $y$ direction, while the perfect-matched layers (PML) are used to truncate the $z$ direction of the simulation domain. The thickness $t$ of slab is chosen to be 100 nm, which is comparable to that of a typical amorphous silicon solar cell. Here in order to see the mechanism behind absorption enhancement patterned thin films more clearly, we use a fictitious material to avoid the strong dependence of optical absorption on material property. The real part $\varepsilon_1$ of the dielectric function is 12 and the absorption length $l_a$ is 3.3 $\mu$m over the entire spectra between 500 and 1100 nm, which corresponds to a wavelength dependent imaginary part ($\varepsilon_2\approx\sqrt{\varepsilon_1}\lambda/2\pi l_a$). Automatic adaptive mesh is used to discretize the simulation domain and the maximum mesh size in the slab is 10 nm. To extract the transmittance and reflectance, a plane Gaussian source is placed 250 nm above the upper surface of the slab and two power monitors are placed above and below the slab to measure the transmitted and reflected power. The reflectance $R$ and transmittance $T$ are obtained by normalizing the recorded power at two monitors by the source power. The absorptance spectra can be calculated with energy conservation $A=1-R-T$. Unless specified, all the results presented below are based on normal light incidence. To quantify the broadband absorption within the entire spectra we are interested in, the integrated absorption (IA) $\eta$ is defined as
\begin{eqnarray}
\eta=\frac{1}{\lambda_{max}-\lambda_{min}}\int_{\lambda_{min}}^{\lambda_{max}} A(\lambda)d\lambda
\label{eta}
\end{eqnarray}
where $\lambda$ is wavelength, subscripts $max$ and $min$ denote the maximum and minimum wavelength we considered in this work. Note that in some literature the ultimate solar energy conversion efficiency is used to quantify the absorption of the structure, where the solar spectra is also considered \cite{Lin_2009, Sturmberg_2012}. Since we are focusing on the mechanism, the simple definition of absorption efficiency in Eq. (\ref{eta}) is employed without considering the solar spectra. All our simulations are carried out using the Lumerical FDTD Solutions package.

\section{Ordered Nanoholes}

\begin{figure}
\centering
\subfigure[Absorption Spectra]{\includegraphics[width=0.4\textwidth]{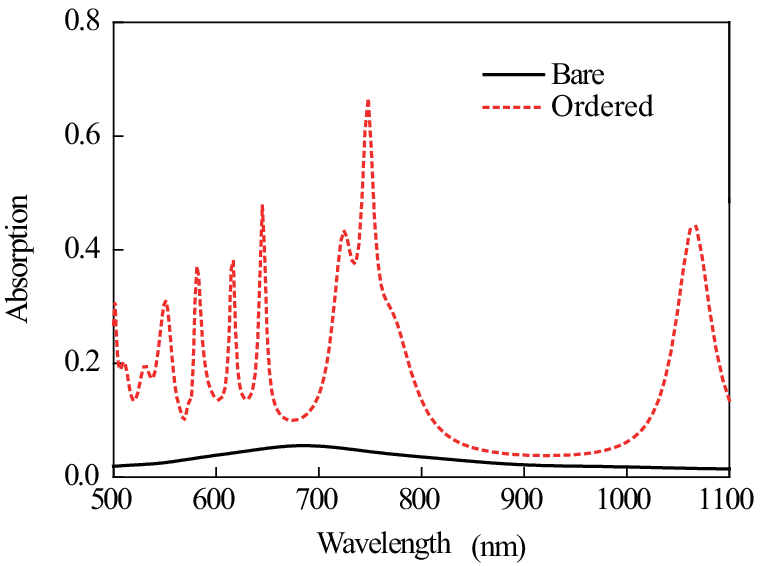}}
\subfigure[Efficiency]{\includegraphics[width=0.4\textwidth]{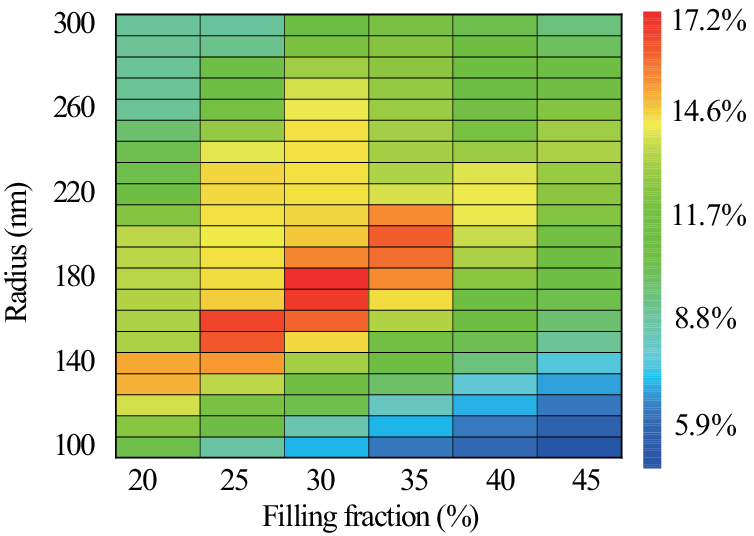}}
\caption{\small (Color online) (a) The absorption spectra of the bare slab, and the slab with an ordered hole array with radius of 180 nm and volume fraction of 30\%. (b) The IA as a function of hole radius and volume fraction. One can see that the optimum efficiency occurs when the volume fraction is around 30\% and hole radius around 180 nm.  } \vspace{3mm} \label{Ordered}
\end{figure}

We first consider the ordered case when the holes are arranged in a periodic square lattice on the slab. This is a typical two-dimensional (2D) photonic slab. Figure {\ref{Ordered}(a) shows a typical absorption spectra of an ordered nanohole array (radius $r=180$ nm) compared to a bare dielectric slab. The bare slab has quite small absorption. The slightly larger absorption when the wavelength is around 700 nm is due to a Fabry-Perot resonance of the slab. The photonic slab with ordered hole array, however, has much larger absorption over the entire wavelength range we are interested in. Several absorption peaks can be attributed to the coupling of incident plane wave with guided resonances induced by the holes, as will be shown later. When these resonances are excited, the electric field within the thin film can be enhanced and the absorption is also enhanced.  We then calculate the IA of different ordered pattern by varying the hole radius and volume fraction, and the results are shown in Fig. \ref{Ordered}(b). The optimized absorption occurs when the volume fraction is about 30\% and the hole radius is 180 nm. It should be expected that there is an optimized volume fraction and hole radius. When the volume fraction of holes are too large, there are too little material to absorb the light. When the hole radius is too large compared to the wavelength, there will be no photonic effect. On the other hand, if the hole radius or volume fraction is too small, the structural feature is similar to a bare slab which has low absorption. The optimized absorption efficiency is about 17.1\%. Compared to the average absorption of 3.1\% for the bare film without through holes, IA of the optimized pattern is increased by a factor of 5.5. With the nanoholes introduced, the optical absorption of the thin film can be greatly enhanced.

\section{Non-uniform radius and random position}

\begin{figure}
\centering
\subfigure[Non-uniform Radius]{\includegraphics[width=0.4\textwidth]{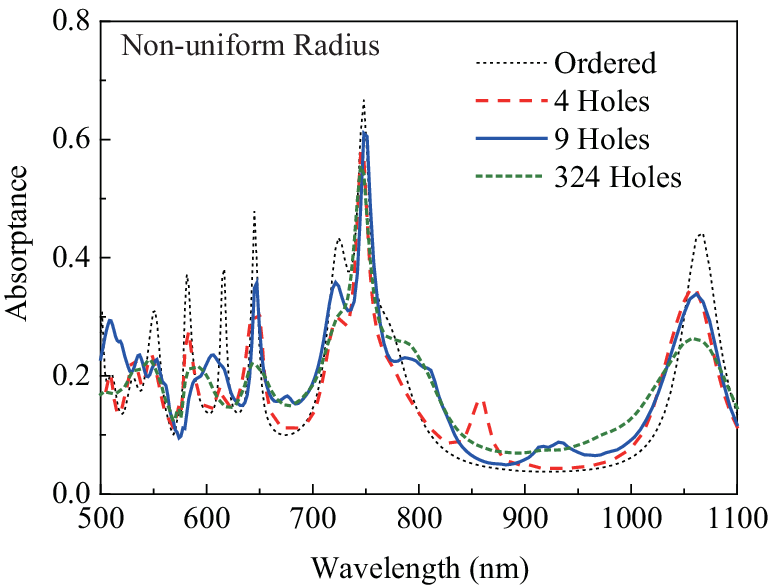}}
\subfigure[Random Position]{\includegraphics[width=0.4\textwidth]{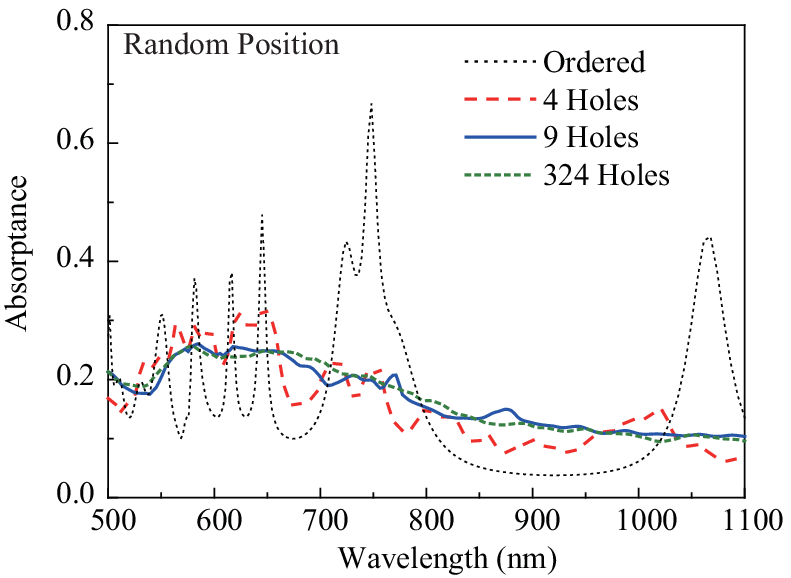}}
\caption{\small (Color online) (a) Absorption spectra of thin film with nanoholes of non-uniform radius (b) Absorption spectra of thin film with random positioned nanoholes. } \vspace{3mm} \label{RD}
\end{figure}


We then investigate the effects of non-uniform radius and random nanohole position. The simulation of non-uniform radius is setup by placing the centers of the nanoholes in a square array, and the radius of each hole is randomly chosen between 120 nm and 240 nm (the average is 180 nm). The randomly positioned nanohole slab has a fixed nanohole radius of 180 nm, but the center positions of the holes are generated randomly. To avoid overlap of holes, a minimum of 20 nm distance between the edges of the holes is enforced during the random number generation process. Since periodic boundary condition is used for our simulation, we are actually generating periodic structures with pseudorandom patterns in a square subdomain. To see the simulation size effect, we considered different numbers of holes in the simulation domain, including 4, 9, and 324 holes. For the case with larger number of holes, the area of the subdomain is also increased to ensure the volume fraction to be 30\% and be comparable to the optimized ordered case. One can see that with a larger number of holes being considered, the pseudorandom structures have smaller degree of translational symmetry and are more closed to a real disordered structure. For the 4, 9, and 324 holes structures, the periods of the subdomain (side length of the square) are approximately 1165 nm, 1747 nm, and 10485 nm, respectively. The absorption spectra of disordered slabs are calculated by averaging 3 different configurations except that the 324 holes case is the  spectra of a single configuration.

The absorption spectra of the slabs containing 4, 9, and 324 holes non-uniform radius hole arrays are compared to that of the uniform radius (ordered) and shown in Fig. \ref{RD}(a). It can be seen that with non-uniform radius, the absorption spectra is only slightly broadened compared to the ordered array. The several absorption peaks for ordered array are also observed in the spectra of slab with non-uniform hole radius. This indicates that the non-uniform radius slab can still preserve the resonances of the ordered array. The IA of the slab with a 9 non-uniform radius holes is 17.6\%, which is slightly larger than that of the ordered array.

Figure \ref{RD}(b) shows the absorption spectra of slabs containing 4, 9, and 324 randomly positioned holes. The spectra of random positioned nanoholes are quite different from that of the ordered array. The absorption spectra of the random position hole array shows a broadband nature and the absorption peaks are completely smoothed out. With larger number of holes, the absorption spectra is smoother. Note that since we apply periodic boundary conditions to the entire simulation domain, the 4 random position holes only create a very small aperiodicity over the ordered case and the long-range order of the system is still preserved. Comparing the 4, 9, and 324 hole cases, the absorption spectra look quite similar, while the spectral peaks gradually broaden as the number of holes becomes larger. The integrated absorption of the 9 hole case is 16.8\%, which is slightly smaller than the ordered case. This result indicates that structural randomness does not always have positive contributions to the absorption spectra.

It is also interesting to compare the nanohole on a slab with the nanowire structure, because  nanoholes on a slab can be seen as a reciprocation of the nanowire array. It was shown that for nanowire arrays,  random position only slightly modulates the absorption spectra, while non-uniform diameter could greatly change the absorption spectra \cite{Bao_2010c, Sturmberg_2012, Lin_2012}.  The distinct behaviors of the two reciprocal structures are likely due to the different types of guided resonance: while in the slab with holes the guided resonances lie in the dielectric region between the holes, in nanowire array the resonance is primarily induced by the leaky mode resonances in the individual wires \cite{Cao_2009}.

\section{Discussions}

\subsection{Random and amorphous pattern}

To see the potential of absorption enhancement by structural disorder, we sampled 100 different cases of the slabs containing 9 randomly positioned holes. The IA efficiencies for the different cases are calculated and the distribution function is shown in Fig. \ref{histogram}(a). The average IA is 16.8\%, which is similar to the 324 hole random position array and only slightly smaller than that of the ordered case. The smallest IA is 14.4\% and the largest is 19.3\%. This indicates that a certain level of aperiodicity does not always enhance the net absorption, in contradiction with the theoretical analysis in Ref. \cite{Yu_2010b}. The contradiction is due to the assumption that the slab has the same photon density of states (PDOS) as a bare slab in Ref. \cite{Yu_2010b}. In our case, the PDOS is modified with the presence of nanoholes. Comparing to the ordered case, the best 9 holes random structure still has 13\% enhancement. Since our ordered structure is already optimized, this absorption enhancement of 13\% is actually quite significant, indicating that further enhancement is possible with optimized pattern.

On the other hand, the amorphous structure has short-range order that could have both improved light trapping and a broad operation band \cite{Vynck_2012}. To compare the performance of random holes with amorphous holes, we also calculated the absorption spectra with amorphous nanohole pattern. The amorphous patterns are generated by the Lubachevsky-Stillinger algorithm\cite{Skoge_2006} on a square area with periodic boundary conditions.  The square has a side length of 1747 nm and contains 9 holes, which is comparable to the 9 hole random case. The growth of hole diameter was stopped at a filling fraction of 70\% and their diameter was set to wanted diameter which corresponds to a final filling fraction of 30\%. To avoid the formation of large ordered arrays, a bidispersity of 0.9 in the size of the holes is used in the generation process. The absorption spectra of one typical amorphous structure is shown in Fig. \ref{Abs_spectra}(b). It shows a broadband absorption in the whole spectral range we consider. The average IA of three amorphous structure is 21.2\%, which is much larger than the ordered hole pattern and the best random pattern we sampled. The amorphous pattern shows quite significant enhancement, indicating that short-range order is important to the broadband absorption.

\begin{figure}
\centering
\subfigure[IA distribution function]{\includegraphics[width=0.4\textwidth]{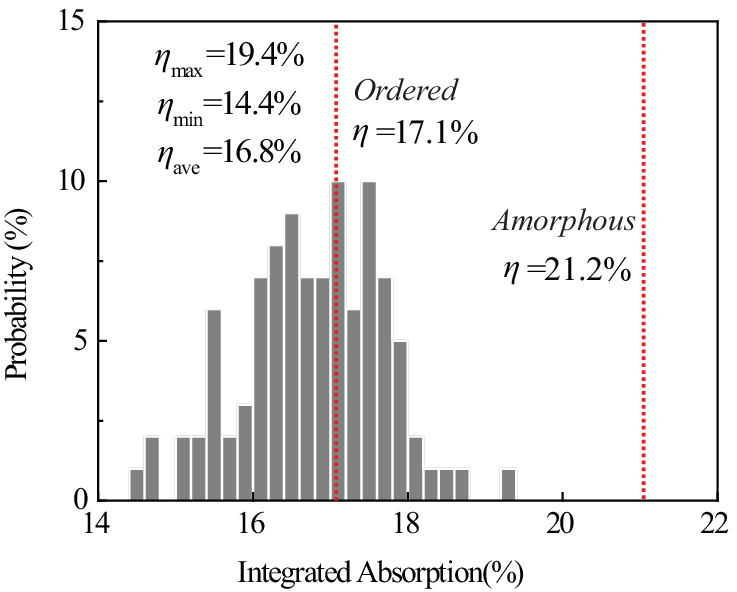}}
\subfigure[Angular dependence]{\includegraphics[width=0.4\textwidth]{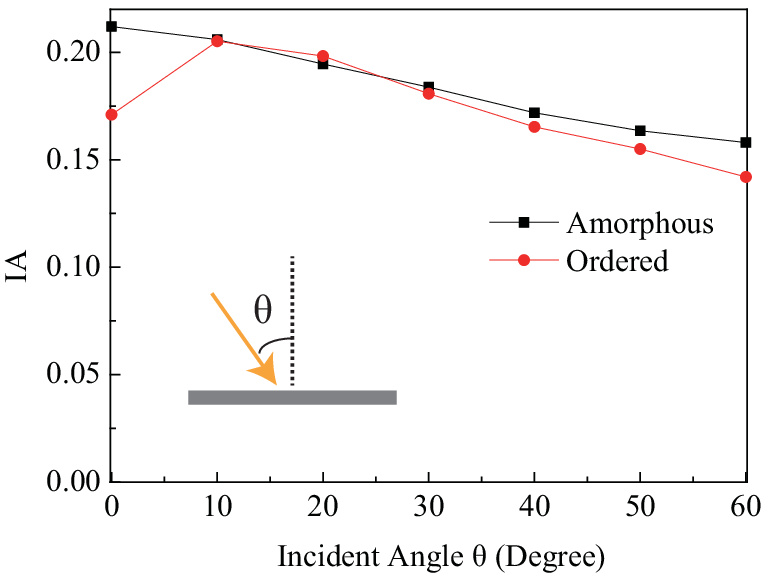}}
\caption{\small (Color online) (a) The statistical distribution of the IA efficiency for the 9 holes random position slabs. The red line indicates 17.1\% and 21.2\%, which are the IA of the ordered and amorphous cases, respectively.  (b) The IAs at different incident angles for ordered and amorphous hole configuration. The IAs are the averaged value for both polarizations, equivalent to non-polarized light incidence. } \vspace{3mm} \label{histogram}
\end{figure}

\subsection{Angular dependence}
Since the solar cell will not be always operating at normal light incidence, we also investigate the angular dependent IAs for the ordered and amorphous hole configuration. Note that for oblique incidence the S and P polarizations are no longer equivalent, so both polarizations have to be considered while doing numerical simulations. The average IAs for both polarizations at different incident angles are shown in Fig. \ref{histogram}(b). It is expected that the IA for larger incident angles should be smaller, since the reflection is generally larger at larger incident angle. The IA for amorphous structure follows this trend completely. However, although ordered structure shows a much smaller IA for the normal incidence, the IAs at small incident angles (10$^\circ$ and 20$^\circ$) increase a lot and are almost the same as the amorphous structure.  The unusual increase of IA at small incident angle is due to the breaking of the reflectional symmetry for oblique incidence. The symmetry of ordered structure prevent some guided resonances in the ordered structure being coupled with plane wave \cite{Yu_2010b}. The oblique incidence breaks the symmetry and light can couple to these modes again. For solar cell applications, the light incident angle can vary. Although the amorphous structure still has better performance than ordered, the advantage is not as much as it looks if only normal incidence is considered. When comparing the ordered and disordered structures, many literature only considered normal incidence \cite{Vynck_2012,Pratesi_2013}. Our results suggest that angular dependence should always be considered while designing the photonic pattern for solar applications.

\subsection{Guided resonance}
\begin{figure}
\centering
\subfigure[Ordered]{\includegraphics[width=0.4\textwidth]{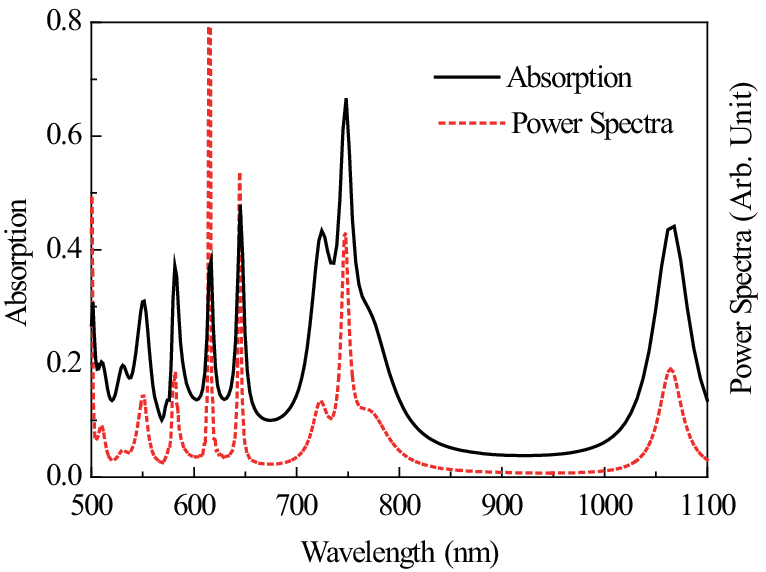}}
\subfigure[Amorphous]{\includegraphics[width=0.4\textwidth]{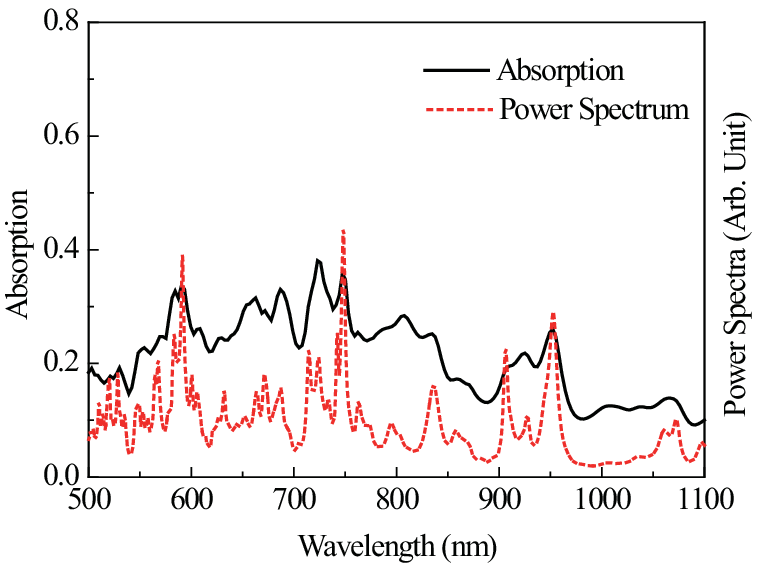}}
\caption{\small (Color online) The absorption spectra and power intensity spectra of the slab with (a) ordered hole array and (b) slab with amorphous hole pattern.} \vspace{3mm} \label{Abs_spectra}
\end{figure}

To find out the origin of the completely different absorption spectra in random position and amorphous configuration, we analyze the resonant modes that can be excited by a plane wave source outside the slab. To do that, we consider a lossless dielectric slab with the same hole pattern as the slab we considered above: one has ordered holes with radius of 180 nm and the other is one amorphous pattern which has 21.2\% IA. A Gaussian pulse centered at 436 THz (686 nm wavelength) from the source plane is used to excite the guided resonances, 30 different observation points are placed within the slab to record the electric field as a function of time. At each monitoring point, Fourier transform is performed to the amplitude of the recorded electric field to obtain the power density spectra, which is also normalized by the source power at frequency domain. The peaks of such spectra reflect the high quality factor guided resonances of the dielectric slab \cite{Fan_2002}. Figure \ref{Abs_spectra} shows the average power density spectra of the 30 observation points together with the absorption spectra. One can clearly see that the peaks of the power density spectra have one-to-one correspondence to the peaks of the absorption spectra, which indicates that the absorption peaks are indeed induced by those guided resonances of the patterned dielectric slab. The disorderness of the amorphous structure reduces the symmetry and provides more resonant modes, but the quality factor of each individual mode is also reduced. These low quality factor modes are desired for better broadband absorption \cite{Yu_2010}.

\subsection{Effect of intrinsic absorption}
\begin{figure}[t]
\centerline{\includegraphics[width=0.4\textwidth]{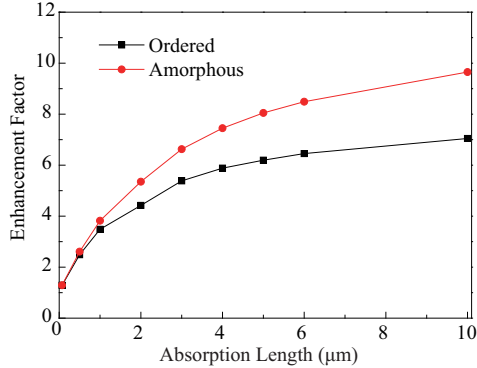}} \caption{\small The IA efficiency of the ordered array and the amorphous pattern when the absorption length varies. }
\vspace{3mm}\label{Abs_Length}
\end{figure}

One important question is whether one can utilize aperiodic nanohole pattern to enhance the overall absorption of ultrathin film solar cells. Based on our analysis, non-uniform radius or aperiodic arrangement on average can slightly enhance the IA absorption. Simple random nanohole position does not enhance IA, but carefully designed aperiodic or amorphous pattern could greatly enhance the absorption. Since the absorption enhancement is due to the guided resonances, a question might arise is that whether the absorption enhancement still presents when there is intrinsic material absorption. To answer this question, we analyze the enhancement factor (IA compared to a slab) when the absorption length of the material changes. As shown in Fig. \ref{Abs_Length}, the enhancement factors increase as the absorption length increases for both ordered and amorphous pattern. If one compares the difference between the ordered and amorphous structure, the enhancement is also larger when the material absorption length is larger. This is because when the absorption length is small, the material is more absorptive and the field enhancement is offset by the loss due to material absorption. Since the enhancement factor of patterned slab is always larger than 1, it is always worth to use patterned slab to enhance optical absorption. From Fig. \ref{Abs_Length}, when the absorption length is as small as 0.5 $\mu$m (approximately equivalent to an imaginary part of dielectric function of 0.66, corresponding to that of silicon at around 500 nm \cite{Palik_1998}), the enhancement of the amorphous pattern over the ordered pattern is merely 4\%. This indicates that the absorption enhancement using aperiodic structure is not important when the absorption of the material is already large. Figure \ref{Abs_Length} can be regarded as \emph{a rough guideline} for determining whether it is worthy to use aperiodic structure for solar absorption enhancement. For example, amorphous silicon is a direct band gap material, which has strong absorption to the photons with energy larger than band gap. Therefore, one can expect the absorption enhancement due to structural disorderness is not significant. In comparison, since crystalline silicon has smaller absorption coefficient in the long wavelength region ($\varepsilon_2<0.66$ for wavelength longer than 450 nm), it is then worthy to utilize disordered structures to enhance the absorption efficiency in the spectral range between 450 and 1100 nm. A final remark is that due to the limited size of simulation domain, we cannot make any conclusion on whether strong localization (or Anderson Localization) of light \cite{Anderson_1958} could play an important role and further enhancement of absorption. This is left for future investigation.

\section{Summary}
In summary, we investigated the optical absorption property of semiconductor thin film with nanohole patterns. The presence of nanohole induces large absorption enhancement compared to the bare thin film, which is attributed to the guided resonance induced by the nanoholes. The non-uniform radius will only slightly broaden the absorption spectra of the ordered array, while the absorption spectra of random position pattern is almost completely broadband. The broadening absorption spectra is due to the additional resonant modes induced by lower structural symmetry. The overall absorption of non-uniform radius and random position hole arrays do not evidently outperform that of the ordered array. If the hole position is arranged in an amorphous pattern which has short range correlation, the overall absorption is significantly larger than that of the ordered array. However, if oblique incidence is considered, the overall absorption of amorphous structure is only slightly better than that of the ordered array. Also, if the intrisic absorption of the material is large, the absorption enhancement using amorphous pattern is not significant.


\end{document}